# Man, machine and work in a digital twin setup: a case study

**Ali Ahmad Malik, Alexander Brem**

University of Southern Denmark, University of Stuttgart

**Pre-print**

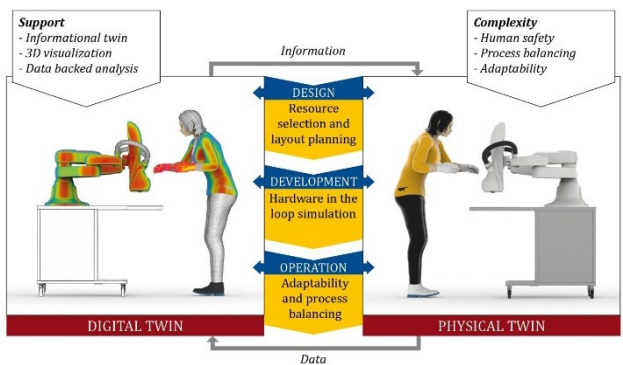
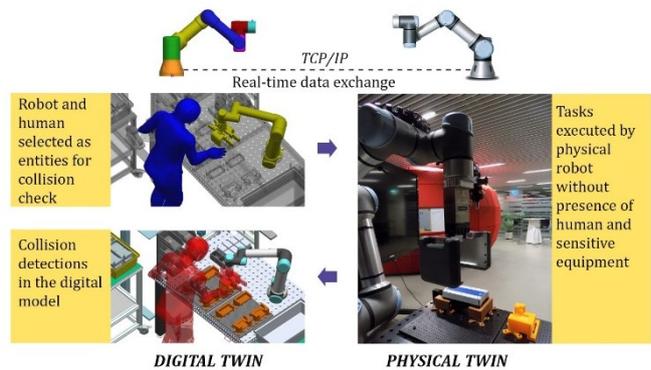
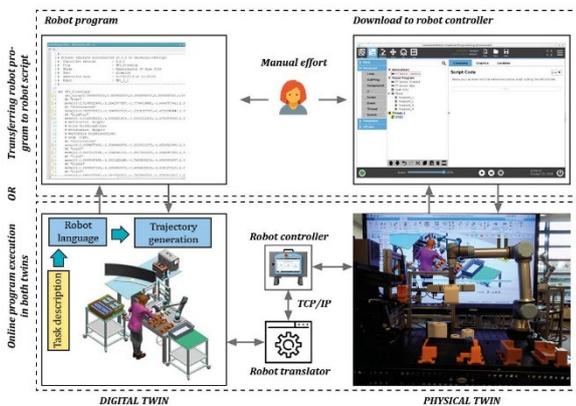
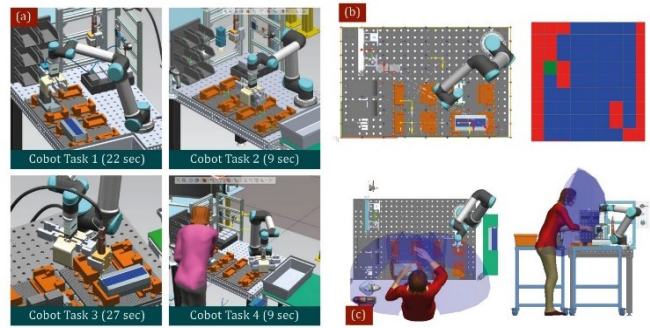

# Man, machine and work in a digital twin setup: a case study

Ali Ahmad Malik[1*], Alexander Brem[2]

[1]Mads Clausen Institute, University of Southern Denmark, Sønderborg, Denmark

[2] Chair of Entrepreneurship in Technology and Digitization, University of Stuttgart, Germany

*E: alimalik@mci.sdu.dk; **E: alexander.brem@ets.uni-stuttgart.de

**Abstract**

*This paper explores the opportunities of using a digital twin to address the complexities of collaborative production systems through an industrial case and a demonstrator. A digital twin, as a virtual counterpart of a physical human-robot assembly system, is built as a 'front-runner' for validation and control through design, build and operation. The forms of digital twins along system's life cycle, the building blocks and potential advantages are presented. Recommendations for future research and practice in the use of digital twins in the field of cobotics are given.*

## 1. Introduction

In the recent decade, a kind of silent revolution has changed the way manufacturing is planned and performed. This leads to a new manufacturing landscape with shorter product life cycles, mass personalization and interactions between humans and robots in close proximity (Rauschnabel et al. 2015) (Mourtzis 2019). The modern manufacturing landscape is getting increasingly complex and production plants are becoming multiplexes of highly complex biomechanical elements.

Similarly, the need for being adaptive, flexible and cost effective is growing. Humans have proven to be flexible production resource but the cost of human effort is continuously increasing (Brem & Wolfram 2014). Human effort can be effectively replaced through automation; however, contemporary automation solutions are not human friendly, and can't coexist with humans (Malik et al. 2019). Due to this reason a large proportion of processes in a manufacturing value chain are still human extensive (Masood & Weston 2011).

If the human and automation can be combined - creating a balance between the flexibility of manual processes and the efficiency and repeatability of machines - the advantages of production flexibility, product mix and reconfiguration can be achieved (Müller et al. 2017). One approach of doing this is using the new generation of industrial robots named collaborative robots or cobots (Malik et al. 2020). Cobots, integrated with advanced safety technologies, are designed for sharing their workspace with humans thus achieving "the right amount" of automation (Müller et al. 2016). However, as the level of collaboration grows, the systems tend to become complex. Due to this reason the industrial application of cobots in labor intensive processes such as assembly is still limited (Salunkhe et al. n.d.).

A way to quickly and safely (re) design and integrates these systems is by developing a virtual space (Choi et al. 2015). Production strategies can be tested and validated in the virtual space before taking them into real practice (Bilberg & Malik 2019). The new Lifecyle approach of virtual modeling is the concept of a digital twin (DT) – a digital representation of a physical system – enabled by the advancement in virtualization, sensing technologies and computing power (Stark et al. 2019).

The paper aims to answer the question: What can a digital twin contribute to address the complexity of a collaborative production system? Based on an industrial assembly case study, this paper describes the development of a digital twin and investigates its usefulness. The case is an assembly problem because assembly is labor-intensive work and has high potential for human-robot collaboration. The case owning company is located in Denmark which is a high-wage economy and is facing the challenges of globalization. Lastly, the case solution aims to use human-robot collaboration where human and robot are sharing the same space and the time, meaning a high degree of collaboration.

## 2. THEORETICAL BACKGROUND

### 2.1. Complexity in manufacturing systems

Systems' theory defines a system (Schlager 1956) as: *'two or more components that combine together to produce from one or more inputs one or more results that could not be obtained from the components individually'.* The components in a system (as individual entities) contain and (their mutual interaction) produce information. The quantity of this information proliferates during the system's life cycle making it increasingly difficult at any phase for an external observer to study or predict its future behavior. The level of difficulty faced by an observer studying a system is referred to as system complexity. However, there is a little agreement on the term complexity and the way it is understood in various contexts. Nevertheless, two important dimensions of a system giving rise to its complexity are the information content and the predictability of interaction-behavior.

Another approach to understanding systems' complexity is the Cynefin diagram (Figure 1) that describes systems in four "phases of knowing" i.e., simple, chaotic, complicated and complex (Snowden 2002). The differentiating property of each of these phases is their behavioral predictivity.

In an manufacturing context the systems' forms of knowing are relevant to the four industrial revolutions. Starting from 1st industrial revolution, the manufacturing systems (e.g. mechanized power looms) were entirely predictable with obvious and transparent inputs, operations and outputs. The 2nd revolution gave rise to chaotic systems where the complexity started to grow but the relationship between a system's elements was turbulent (e.g. an assembly line). Complicated manufacturing systems, with a network connected components are the result of 3rd industrial revolution. Networked components are multi-component structures but the connection between the components is linear and straight forward. As the adoption of computers in manufacturing actives surfaced the behavior of the manufacturing systems became largely predictable. An example of a complicated system is an automotive assembly line of the modern era.

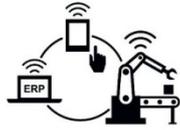

Figure 1: Complexity and phases of knowing in the evolution of manufacturing systems.

Whereas complex systems are a large network of components with many to many communications and with sophisticated information processing that makes the behavior of the system difficult to predict. Complex systems have a property of surprise, as in 'I didn't see that coming', and such a surprise can even be fatal (Grieves & Vickers 2017a).

With growing automation and many to many relationships, the complexity of predicting the behavior is growing. It is furthermore important if the 'surprise' can result in an accident, especially if the system involves interaction with humans. There are other elements of time, energy and effort that can result in economic wastes if the system fails.

### 2.2. Complexity in human-robot collaborative systems

Humans are complex. Not only as a group of humans, but even as individuals, they exhibit a complex behavior which is often difficult to predict (Snowden 2002). The interaction of humans with another system (existing at any phase of knowing) can escalate the total information content and fabricate unpredictability, creating some sort of 'complexity'. This is evident from Perrow's work, 'Normal Accidents' (Perrow 2011) on describing the inherent danger of complex systems and pointed to a common thread of 'human-element' when they are interacting with complex systems (Grieves & Vickers 2017a).

With the rise of Industry 4.0 technological trends, a growing trend in manufacturing is automation with human-friendly robots or *cobots* (Figure 2). A cobot is a robot designed to be safe when working alongside humans. Since humans are prone to errors, human-robot systems do not always perform flawlessly. From a system's perspective, the complexity of an HRC variant oriented production system can be described in four different dimensions i.e., many components, non-linearity, relationships between components and reconfiguration. An answer to the complex behavior of an HRC system is to somehow (re) design the system in retrospect, predict the future under maximum known variables and implement it.

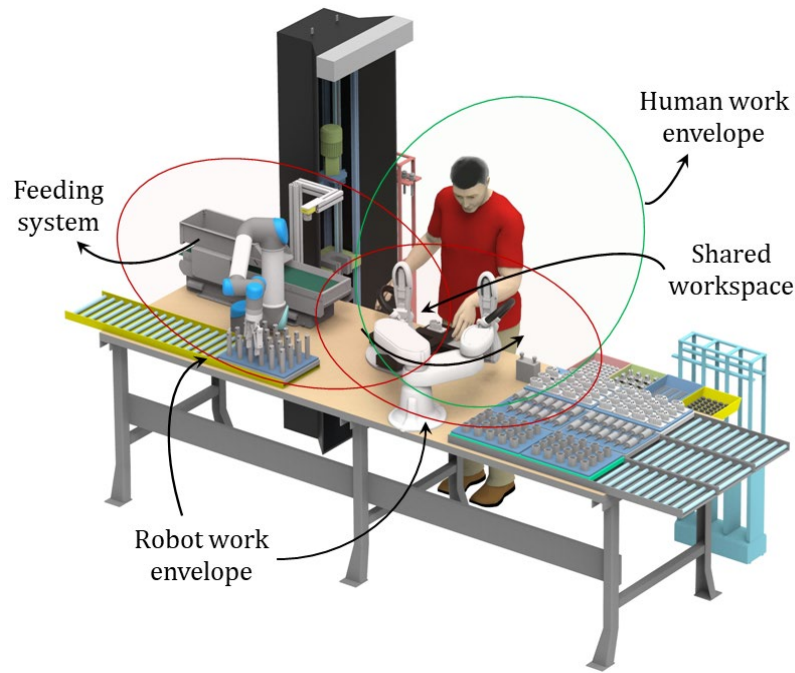

Figure 2: Human-robot collaborative workstation.

### 2.3. Digital twins to address complexity

A digital twin (DT) is a digital representation of both the elements and dynamics of a physical system; enabled by the advancement in virtualization, sensing technologies and computing power (Stark et al. 2019). It mirrors the real-time operating conditions of a physical system (Rosen et al. 2015). The concept of DT emphasizes utilization of digital model of a physical system by linking each virtual element to a corresponding physical asset. The digital twin becomes a mirror and can act as a 'front-runner' to the future behavior of its associated physical twin or system.

The concept of DT evolved from a hardware twin that was developed in NASA's Apollo program consisted of two identical space vehicles. During the space mission, one vehicle remained on the ground while the second went up-to the space. The ground vehicle was continuously mirroring the flight conditions of the in-orbit vehicle to enable ground experts to better assist astronauts in orbit (Schleich et al. 2017).

A different approach to the concept of DT as a "Conceptual Ideal for PLM" (PLM = Product Lifecyle Management) was presented at University of Michigan (Kahlen et al. 2016). The approach assumed that every system can be considered as a subset of two systems i.e. the physical system existing in the physical world, and a virtual system existing in virtual space and containing all (necessary) information about the physical system. The bi-directional relation between physical system and digital system can make insights into a system's performance improving product design, manufacturing and service throughout the system's life cycle (Schleich et al. 2017).

The basic idea of having an (informational) virtual twin to understand the complexity of a physical system is not new. Engineering design and build problems have always been solved through an informational twin. In earlier times, when developing a system, the virtual twin was an image in human-mind though limited in its ability to answer questions about system's performance. Only in the mid-20th century, it became possible to draw a virtual model as a 2-dimensional (2D) CAD (computer aided design) object and then as 3D models, and later as dynamic simulations (Figure 3). During development of a system, these virtual models are created early as the first form containing maximum details of the proposed physical system. However, in practice, these virtual models get obsolete once the system has evolved into its operational form. The reason is the effort required to make the digital model intelligent to make sense of the operational behavior and comparing with its algorithmic solution and assisting in day to day production variations.

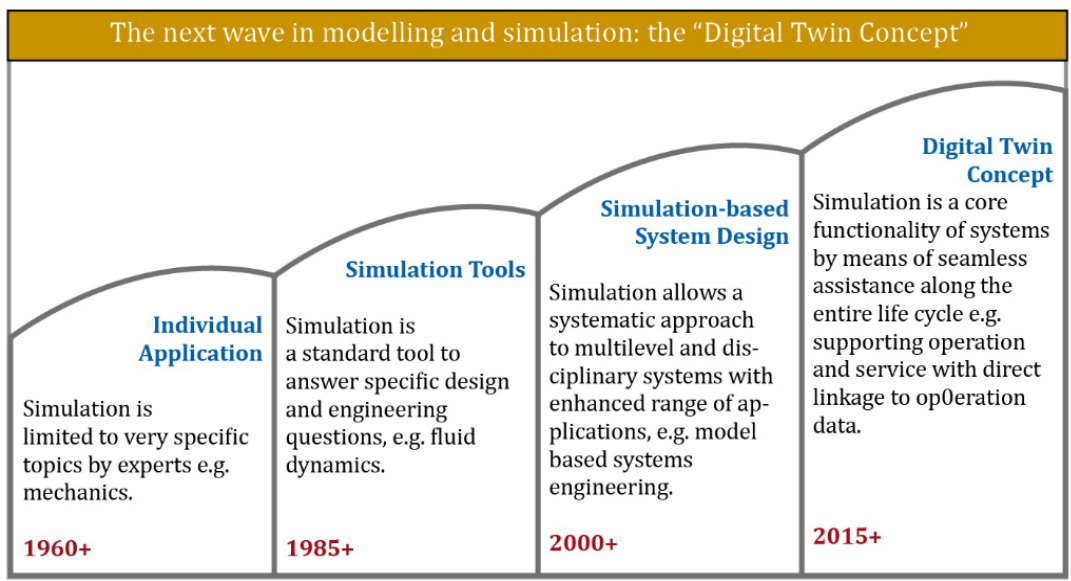

Figure 3: Use of CAD model during system development.

With hyperscale computing, commoditization of sensor equipment, augmented/virtual reality and data analytics platforms, it has become possible to develop time dependent three-dimensional models of complex physical systems. It extends the usefulness of virtual models developed in the design phase to the lifecycle of a system.

## 2.4. Related work

A multitude of definitions and understandings of digital twin conception are available in literature, however the core idea hovers around a synchronized multi-purpose simulation of a physical system that (through sensor updates etc.) mirrors the life of its physical twin (Qi et al. 2019) (Wilhelm et al. 2020). Although the literature displays application of DT in various contexts such as aircraft design (Tuegel et al. 2011), buildings (Naticchia et al. 2019), healthcare (Liu et al. 2019), and farm management (Verdouw & Kruize 2017) etc.; but an overwhelming interest has been observed from manufacturing applications. Several contributions have been made in this regard such as digital twins for product design and service (Tao et al. 2017), digital twin shop floor (Tao & Zhang 2017), DTs for factory design (Guo et al. 2019), production line performance (Fera et al. 2020) etc. Nevertheless, most research is available in form of conceptual models and simulation but application to real-world scenarios is limited.

The keynote paper (Wang et al. 2019) of the CIRP (Society of Production Engineering) pointed towards the importance of DTs for HRC production systems. It was presented that a DT can simplify combining and aligning the function, structure and behavior of an HRC cell with symbiotic interplay of the humans in virtual models. Grieves (Grieves 2019) (among the earliest researchers in the field of DT) has also presented the potential value of the field of cobotics by using DT.

A digital twin integrating human-robot collaboration with level of autonomy for container unloading was presented by (Wilhelm et al. 2020). The authors discussed the combination of adaptive automation and digital twin enabling an operator to remotely control an autonomous system if needed.

The significance of digital twins for HRC assembly systems has been discussed by (Bilberg & Malik 2019) where the relevance of DT in relation to human-robot collaboration was discussed. It was presented that DTs can support HRC systems however, the study remained limited only to the HRC challenges in operational phase without a lifecycle approach. Various aspects of achieving a digital twin such as forms of DTs along system lifecycle, building blocks of the DT and demonstration of the discussed concepts were also missing.

The life cycle context of production systems and their virtual modeling has been presented by Kovsturiak (Kovsturiak & Gregor 1999). The study made recommendations for effective utilization of discrete-event simulation for design, operation and continuous improvement of complex manufacturing and logistics systems. Kibira (Kibira et al. 2015) presented different methods and tools for simulation application in life cycle of production plants. It was argued that traditional simulation models built for concept development and design are different from models for operational decision-making but the distinction is getting blurred with smart manufacturing systems.

To cover the gaps, the present study is contributing by taking a holistic life cycle approach of digital twins for human-robot assembly systems. The building blocks and various forms of DT are presented. A physical demonstrator of an industrial assembly case is developed together with its DT to substantiate the presented concepts.

# 3. DIGITAL TWIN DRIVEN HRC SYSTEM

A digital twin model for an HRC production system is shown in Figure 4. The digital twin system is composed of two interconnected spaces i.e. physical and digital. The digital space is a three-dimensional virtual representation of the coexisting or collaborating human and robot while the physical space is the real production system composed of humans, robots and other hardware. Each element in virtual space is displaying the design and operating behavior of a connected physical object in the physical space.

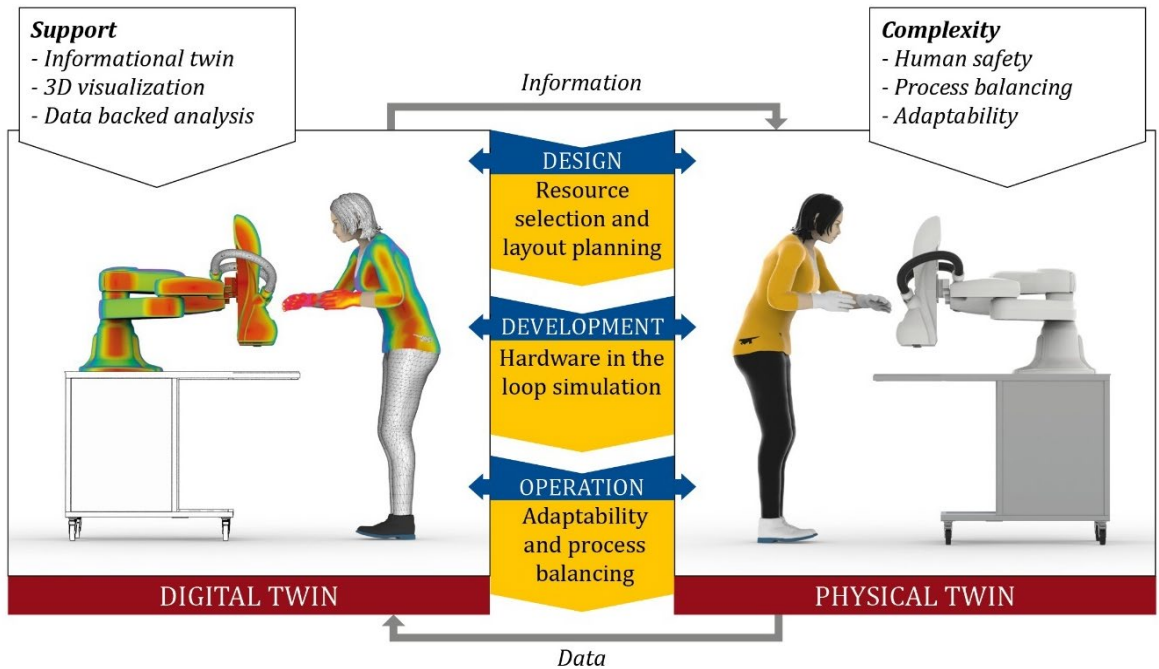

Figure 4: Digital twin in an HRC system.

'Systems engineering' deals with the whole lifecycle of a (complex) system i.e. from analysis and design, through implementation and operation to its redesign and reconfiguration (Kovsturiak & Gregor 1999). The activities in design, implementation and operation must be structured and iterative (Malik & Bilberg 2017). It is proposed that when developing an HRC production system, its digital twin be developed as early as possible during the idea generation and must evolve in parallel to the physical twin. As the development evolves, both the DT and PT must continuously be updated as a reflection of each other. As with all twins, there is a first born. In the case of digital twins of production systems, the first born is the virtual twin because the idea, shape, functionality always precedes the actual realization of the physical form of the product. An HRC production system designed with a DT approach can offer:

- Fast integration,

- Production reconfiguration, and

- Safety

To make DT and PT useful for each other, a DT assessment-model is suggested. The assessment model is composed of sensing the data, evaluating the data, defining non-compliance, setting up a solution, simulating the solution, forming a strategy and implementing it. The assessment model is usable for HRC system's life cycle to optimize system performance.

The below sections describe phases, building blocks and application of DT along the life cycle of an HRC production system.

## 3.1. Forms and phases of an HRC digital twin

By integrating DT into systems engineering for HRC, the physical system can have its informational twin all along its life cycle. The informational twin or DT co-evolves with the physical system and takes different forms corresponding to level of details, connectivity with the real-world data and its synchronization with the rest of the systems. Thereby a classification of DTs can be established (Figure 5).

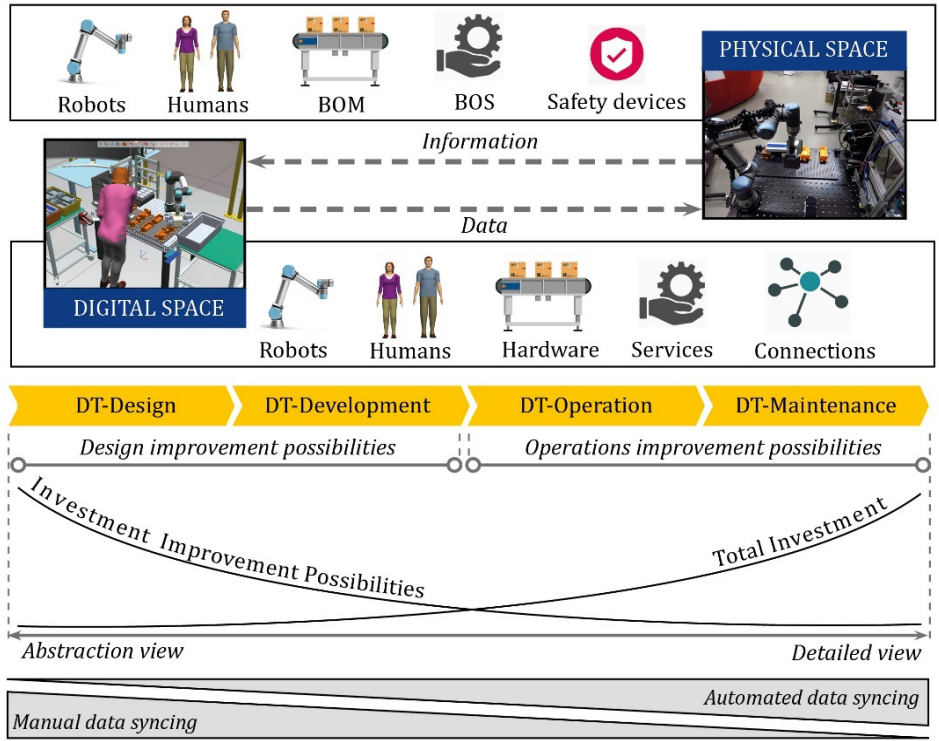

Figure 5: Forms and phases of a digital twin.

Two forms of DTs have been discussed in the literature (Grieves & Vickers 2017b) (Qi et al. 2019); (a) DT that contains static and dynamic informational sets required to describe and produce the physical entity; and (b) a DT where different information sets (monitoring, sensing, service etc.) are integrated for ubiquitous tracking. However, a DT can have additional forms and with a systems engineering approach, these can be better described depending on the life cycle phase of its physical twin.

### 3.1.1. Digital twin – Design

When developing a new physical system, a digital twin is often developed earlier than the physical twin. It can help to generate and validate the initial design, behavior, layout etc. Although, at the design phase, the corresponding physical twin is not existing, but the DT-Design is still referring to a hypothetical future physical twin. Even without a real-time connectivity, the digital twin of the production system can make it possible to experiment several what-if scenarios for achieving faster, safer and better design. The parameters to achieve an effective HRC production system may include selection of robot arm, workstation design, layout, fixtures and economic evaluations.

### 3.1.2. Digital twin – Development

Moving towards DT-Development, results from the DT-Design are utilized to develop elements of the physical twin corresponding to their counterparts in the digital twin. An HRC system may involve development of the workstation, fixtures, feeding devices and other hardware. Bill of materials (BOM) and bill of processes (BOP) are generated and the physical system is developed accordingly.

### 3.1.3. Digital twin – Commissioning

In this phase, initial connectivity between the physical and the digital twins can be established by connecting the DT to a real controller or PLCs to detect potential errors. This method is similar to virtual commissioning (VC) (Lechler et al. 2019). VC or hardware-in-the-loop simulations can help in reduction of development time by virtual testing and integration well before the real commissioning (Lee & Park 2014). The physical robot in PT can be connected with the virtual robot in DT enabling the physical robot to perform the tasks as designed in the DT.

### 3.1.4. Digital twin – Operation

The DT is extended for real-time communication with the physical system during its operation for behavioral analysis and performance optimization. The interlinking of production planning and control databases can support dynamic scheduling of the production orders and allocation of workforce. The DT can simplify the reconfiguration or repurposing of the production system under demand fluctuations.

In DT-Design the assessment is done manually. However, in DT-Operation, the real-world data is synchronized with the DT and the assessment cycle can become automated. Human effort is still needed when a solution to the given situation is generated and is presented to the human user before allowing its implementation.

### 3.1.5. Digital twin – Maintenance

Maintenance is an integral part of the production system. Mixed reality technologies such as augmented reality (AR) or chatbots can be integrated with a DT to enable maintenance personals for maintenance, fault detection and training.

## 3.2. Building blocks of a digital twin system

### 3.2.1. The digital space

As suggested by Tao (Tao et al. 2017) the DT is built in four layers i.e. geometry (creation of 3D CAD objects), physics (kinematics of robots and human), behavior (placement of CAD objects in the scene), and rule (assembly process sequence). Since each element of a digital twin corresponds to its counterpart element in the physical twin the modelling of digital space consists of identical activities as a physical space (Figure 6). Although the digital space is an integral element of the DT system however, a standalone digital space without an interlinking with a physical twin is incomplete. It is important that a DT needs to simplify the complexity and create value for its users. When modelling a DT, not all elements or aspects of the PT need to be modelled. Because information retrieval has a cost and is a tradeoff with time, energy and effort. For example, the rotation of the screwdriver can be modelled and tracked in the DT to know real-time torque, and speed. However, if in the given case, this information is not creating any value then it is of no use and is only growing the complexity.

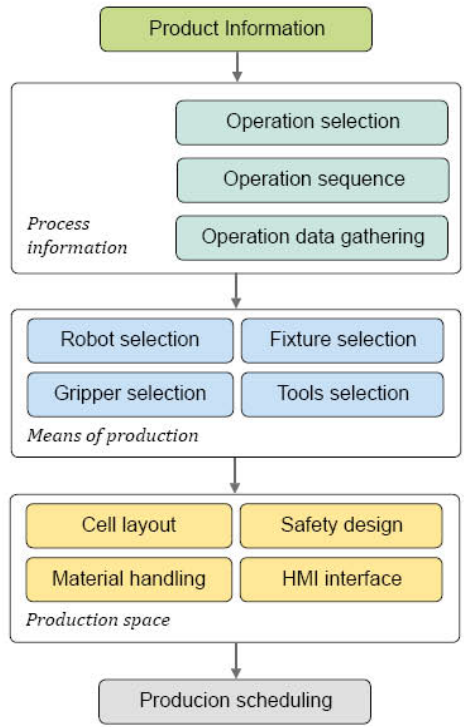

Figure 6: The structured process of designing a production system (adapted from (Chavarr'ia-Barrientos et al. 2018)).

The digital space is composed of different elements such as:

a) **DT Environment**

The digital twin environment is the space where the digital twin is created, visualized and interacted (e.g. on a computer screen). It is a 3D space where the virtual models of the production systems are imported or created. The environment must be able to sync data with other relevant software e.g. CAD modelers, statistical analysis tool, data loggers etc.

b) **DT of assembly components**

A holistic DT of an HRC system begins with the digital modeling of the components being assembled in the system. It may be desired to redesign the components for ease of automated assembly but that is out of scope of this study. This study assumes no change in the design of the product and virtual models of the assembly components are necessary to holistically model the DT processes.

### c) DT of production resources

The digital model of each of these can be (1) supplied by the vendor; (2) offered in the library for standard industrial equipment; (3) created via intensive scans of the part reconstructed into a CAD model; or (4) designed and modeled by the designer (especially for the custom designed components and sub-systems). The digital elements are imported into the DT environment and are placed according to the production requirements.

Digital models of robotic manipulators are often provided by the robot manufacturer. The kinematics, joint limits and maximum joint speeds are defined. The dynamic behavior and constraints of other equipment such as, grippers are defined similarly. Visually accurate 3D digital mannequins can be developed to optimize human activities in computer. The digital human models are often based on databases of human body shapes such as ANSUR II (Anthropometric survey of US army personnel) database (Gordon et al. 2014).

### d) DT of processes and services

A static digital model of the production system is conceived from the previous step. The next step is to define the dynamic behavior of human and robot to execute the tasks. The performance of robotic and human actions needs to be continuously tracked; therefore, continuous simulation is used at this stage.

#### 3.2.2. The physical space

A DT always refers to a physical space. Although at the design phase, for example, there is no physical space existing but the DT is still referring to a hypothetical future physical space which is driving the evolution of the digital space. The physical twin is conceived once the design validation is completed in the virtual models.

#### 3.2.3. DT connections

The connection between the physical space and the digital space is established. The connection or the data syncing can both be manual or automated. However, at the operational phase the data communication must be automated.

#### 3.2.4. DT data

It is the data repository to save the data generated by the digital twin, the physical twin and their interlinking.

### 3.3. Usability of a digital twin for HRC production system

The following section provides an overview of the domains where the digital twin can help in systems engineering of an HRC production system. The usefulness is described in three phases of design, integration and operation.

### 3.3.1. The design phase

Selection and optimal placement of production resources is important in a production setup to avoid wastes. The co-existence of humans and robots imposes further safety implications when planning the layout. In this regard, the following experiments can help to achieve a safe and waste free layout:

**a) Reach and placement evaluation**

The selected robot (under its joint limits) has a certain allowable reach distance. The placement locations of the robot as well as the equipment will determine if the robot can safely reach the desired locations within its workspace. These tests can be performed in a simulated environment. Similarly the reachability of human arms (depending upon its body measurements) without bending the body can be examined. The goal is to have minimum cycle time, minimum collisions and safer working conditions.

**b) Collision tests**

In an HRC, where humans are interacting with cobots and frequent production changes may be desired, altering the human-robot trajectories to move from one point to other. Collisions are likely to occur in such a dynamic environment. Although cobots are designed for workspace where a collision between human and robot is likely to happen, however frequent collisions reduce productivity. Efforts must be made to design the robot trajectories encountering minimum collisions with the human.

**c) Ergonomic analysis**

Manual work is associated with various human performance concerns including work postures, load on human body during weight lift and task frequency. A large variety of these ergonomic issues of production systems can be simulated and tested early in the design phase (Chaffin 2008).

### 3.3.2. The integration phase

By having a mix of physical and digital resources a mixed reality environment can be enabled. For example, a physical robot can be connected with a virtual robot (available in the DT model) to study its behavior. When connected, the virtual robot will follow the trajectory of the physical robot. If the physical robot is in an empty area, the virtual model can be populated with tools, equipment and humans to create the future scene of the physical space. Thus, a task executed at the physical robot enables safe analysis in the virtual space. Similarly, the signals and cycle times can be tested with the use of PLCs and defined PLC logics.

### 3.3.3. The operation phase

**a) Dynamic task allocation**

To complete the assembly, humans and robots both have some unique as well as common capabilities. To choose the right resource to perform a task can be based on the skills of the humans and the robots (Müller et al. 2017) (Ranz et al. 2017). Various other factors can also influence the suitability of a task for the robot e.g. physical and geometrical properties of the components, assembly characteristics, safety implications, feeding and joining method involved. Method to quantify the automation complexity of assembly tasks has been discussed by (Samy & ElMaraghy 2010) which takes care of assembly attributes contributing to automation complexity and a score as a representation for automation potential can be generated. A decision-making arena can be formed where the tasks can be classified as robotic, manual or either robotic or manual. The DT model can estimate the cycle time for each task considering the key positions (pick and place) for the robot and generating robot trajectory. A final assembly plan can be developed depending upon task sequence, minimum cycle time and avoiding any idle time.

### b) Robot control program

A great deal of human effort in robotic applications is associated with the programming of the robots (Rossano et al. 2013). Even in case of cobots – with easier programming interfaces as compared to conventional industrial robots - the programming is still a time consuming and tedious task for industrial applications. With the use of a DT, the robot program is intuitively generated from the DT environment. Once the desired operation is tested virtually the robot program is transferred to the connected cobot that starts working as the robot in the digital twin.

An online connection between physical and virtual robot can dynamically transmit any movement made in physical space to the robot in the virtual space and vice versa (Figure 7). Thus, avoiding any need to have any additional programing. Since the simulation has robotic functions in a robot understandable language, the robot program can be transferred back to the simulation to make simulation run according to the physical robot for any verifications.

### c) Data logs and performance analytics

real-time performance metrics, optimization analytics and alerts for system performance can be recorded in a data repository for performance evaluations. Data connectivity in the operation phase is needed to feed in real world data and making evaluations under continuously changing variables.

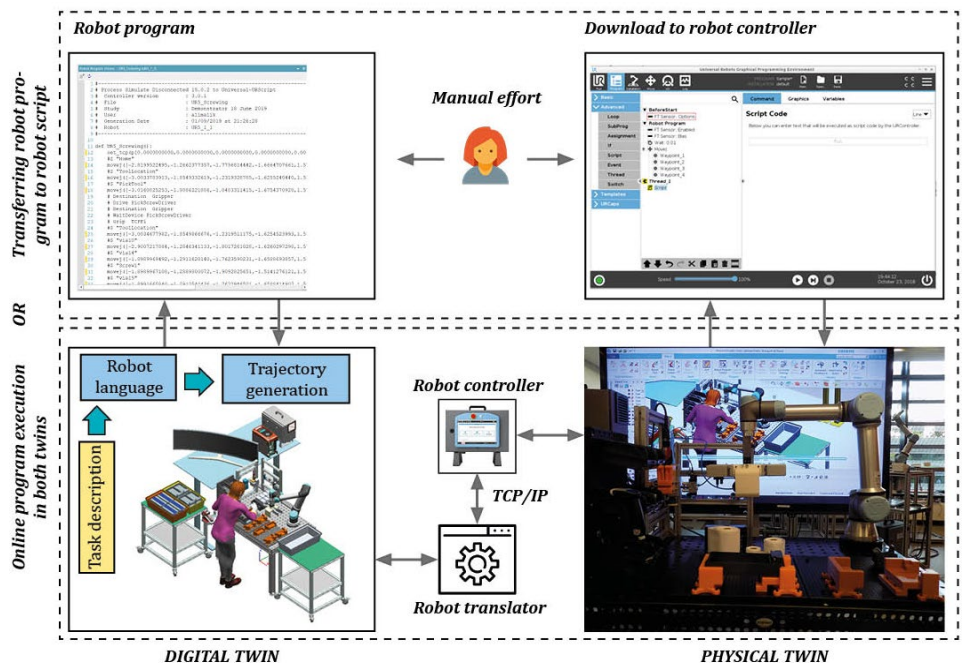

Figure 7: The robot program in physical and digital twin.

# 4. AN INDUSTRIAL CASE

## 4.1. Problem description

The case used to demonstrate the development of an HRC digital twin is from a manufacturing company producing battery packs (Figure 8). The battery pack, in the given case, is used to power electronic linear actuators used in hospital beds for linear movement. The battery-pack consists of eight unique parts; corresponding to eight assembly tasks (each component corresponds to one task). The conventional mode of production was manual where two human operators were completing the assembly. The HRC cell is supposed to reduce man hours but achieving the same production volume. The reduced man hours will result in reduction of production cost and relieving man hours from repetitive tasks.

Each assembly task is evaluated for its ease of HRC automation. Task evaluation for cobot automation is different from conventional robotic automation as additional parameters, particularly for safety implications, need to be considered. A complexity-based task allocation method (Malik & Bilberg 2019) can be used to decompose each task into its attributes identifying the tasks with a higher potential for automation. A total of 4 tasks are selected to be designated to the robot. The remaining 4 tasks are kept manual.

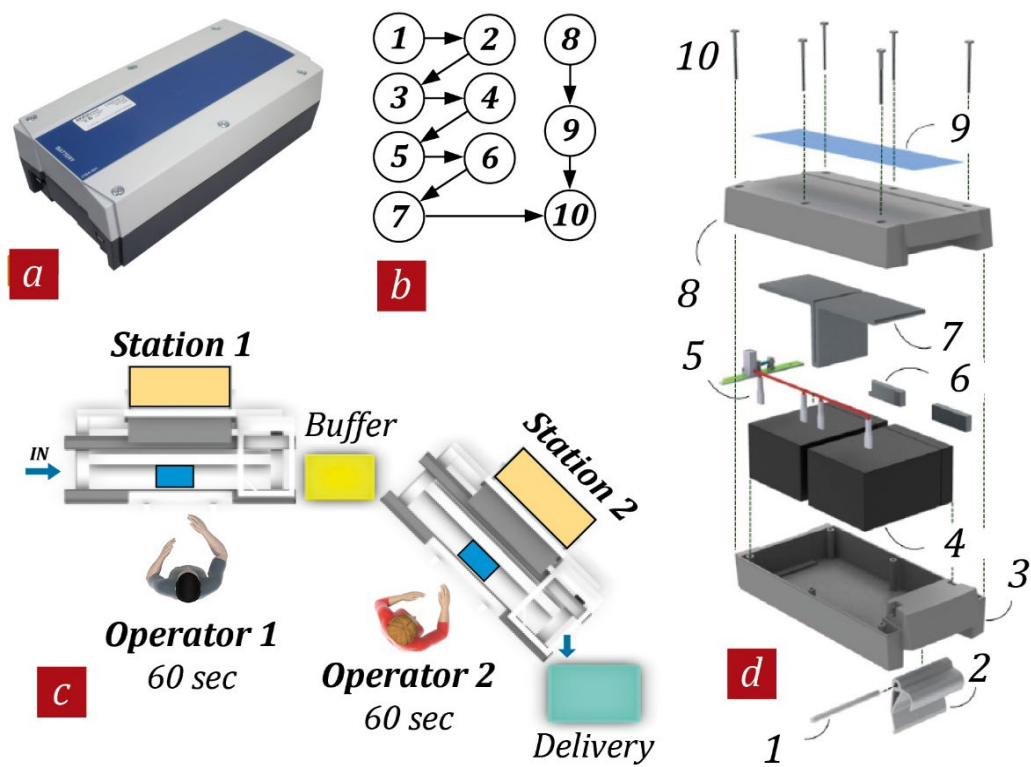

Figure 8: Assembly components considered for HRC automation and the manual cell being used in the case company.

## 4.2. Building the digital twin

The digital twin (digital space in the DT model) is a three-dimensional dynamic environment composed in Tecnomatix Process Simulate (TPS) software (Figure 9). The proposed HRC assembly station comprises of a robot manipulator and a human which complete the assembly in a collaborative fashion. The suggested robot manipulator is a Universal Robot UR-5 e-series. The robot has 6 degrees of freedom, a payload capacity of 5 kg and a reach of 850 mm. The robot is equipped with a parallel SCHUNK gripper EGP 64-N-N-B with a finger length of 40 mm. The gripper in its default design is not enough to handle the large parts (>120mm) used in the case study. Therefore, extended fingers are designed to be produced with additive manufacturing.

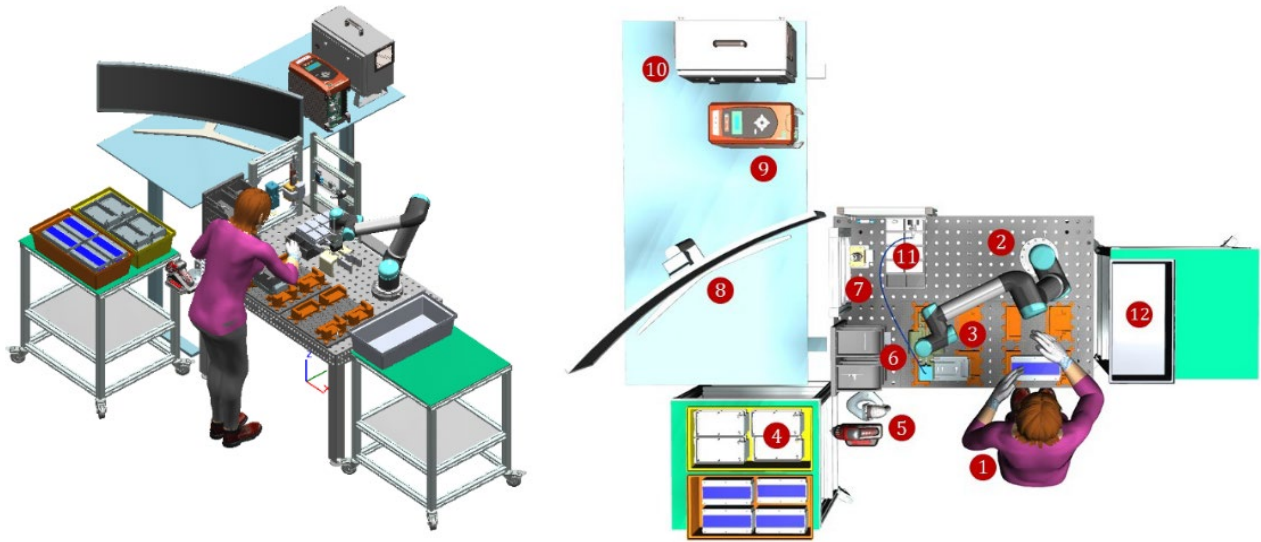

Figure 9: Digital twin model of the HRC system.

Computer aided design (CAD) model of the cobot is received from the robot manufacturer. It is imported into the simulation environment and its upper/lower joints limits are defined with acceleration as defined in robot manual. Other hardware resources are created in Autodesk Inventor as 3D models and then imported into TPS environment. TPS can import CAD data in JT (Jupiter Tessellation) format. A digital human mannequin of the human worker is integrated to conduct ergonomic evaluations. Since the case-factory is located in Denmark, an average size of human population is selected from Danish statistics report 2017 (Denmark 2017). The selected human corresponds to a height of 182cm; BMI < 25 and waist to hip ratio (for females).

Virtual sensors are also integrated to emulate physical sensors to detect the presence or position of components. The sensors used in the virtual model are listed in Table 2 while their locations are shown in Figure 9. A transition logic is shown in Figure 10 that shows the arrival of product to station B executes the robotic operation if the robot is at *Home* checked through joint value sensor. If all the logic is fulfilled, the operation 2 is executed.

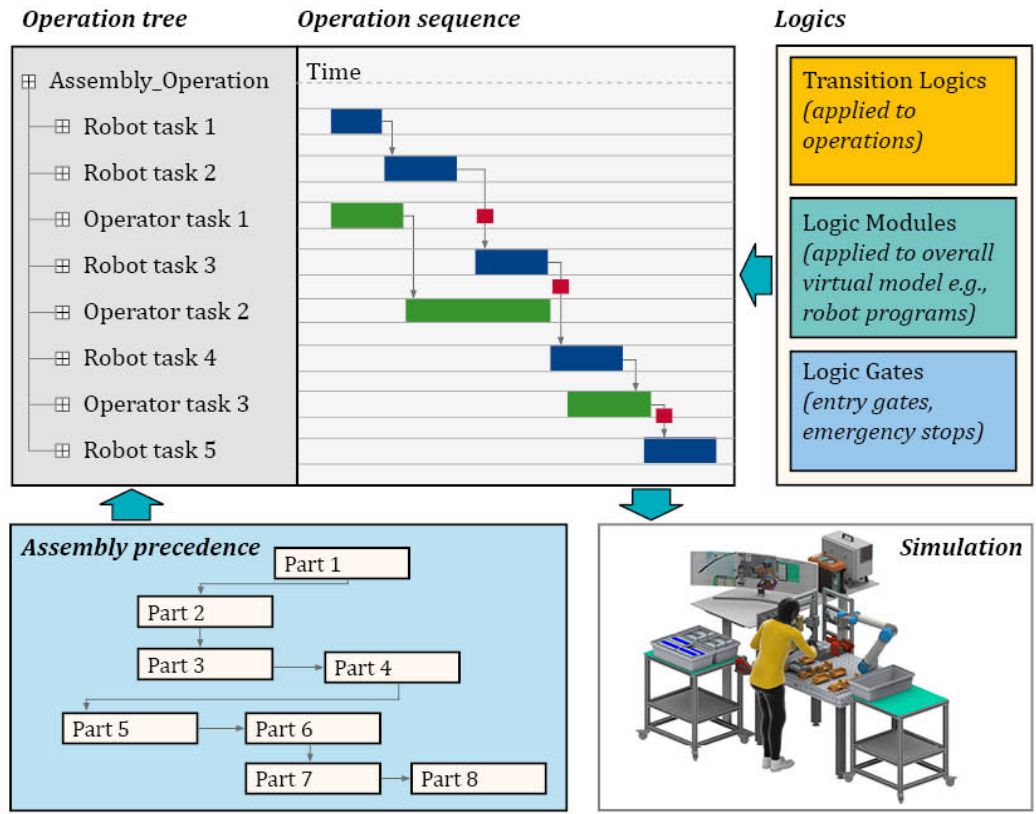

Figure 10: Assembly tasks in the logic driven simulation.

Table 2: Task allocation and process balancing of the HRC assembly cell.

| Sr Nr. | Sensor type | Description |
|---|---|---|
| 1 | Proximity | The sensor is activated when an object enters with a specified distance from a selected resource |
| 2 | Photoelectric | The sensor is activated when an object crosses the path of a defined beam emitted from the sensor |
| 3 | Property | Is activated when an object with a certain property enters in a specified range |
| 4 | Joint value | Sensor to obtain online feedback of a specified joint |
| 5 | Joint distance | A sensor linked to the joint of a device or robot doe its detection range |

### 4.3. Tests and analysis

Once a digital twin is achieved for the proposed physical HRC assembly system, the tests are performed to deal with different issues of system design. An effective HRC layout must ensure maximum utilization of resources, minimum movements, least travel time from one location to next and the safety of human operators.

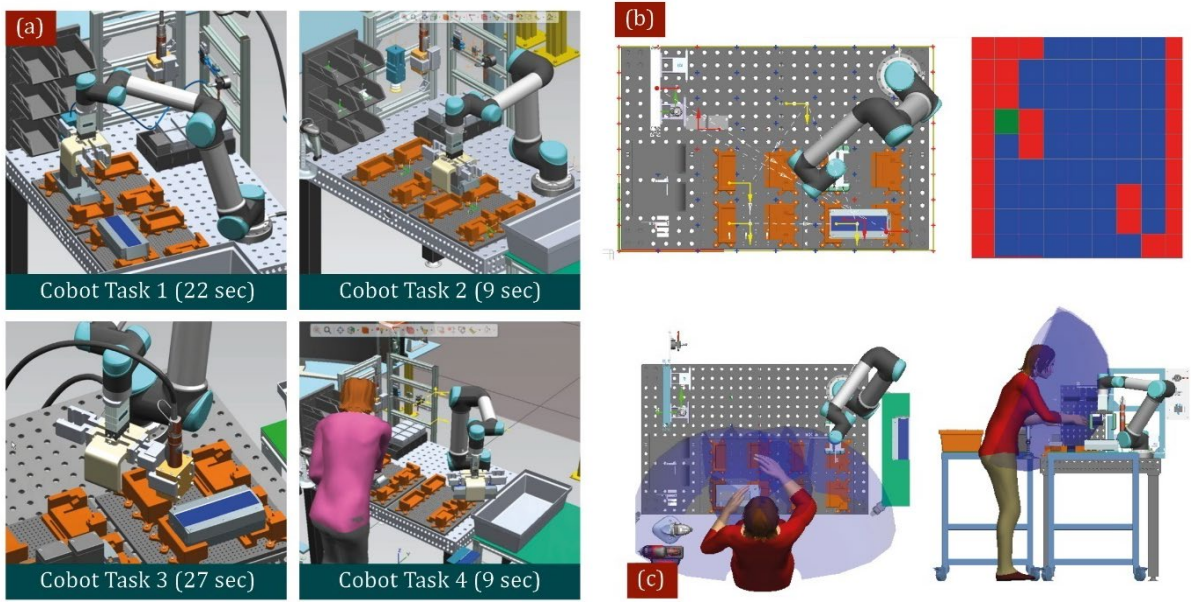

Figure 11: (a) Cycle time estimation; (b) Blue boxes denote areas where the robot can be placed given the reach location while red areas are places where robot (if placed) can't reach to the point of operations; (c) The grasp envelops of the human operator. The envelope shows the areas where the human can reach without bending the body.

The reach test of the robot arm - with respect to its placement location - displayed the points where the robot can reach and components and fixture can be placed (Figure 11). Robot swept volume (Figure 12) identified the space occupied by the robot during its operation. Minimum human activity is ensured in the robot swept volume. Similarly, a human envelope (given its body dimensions) identified the space that human can safely reach. A vision-window enabled the scene to be viewed through the eyes of the human mannequin. Considering the findings, a layout is developed and tested in the simulation environment.

Each robotic and manual task is performed in the DT. A logic is defined that ensures the completion each task and initiates the next task. Finally, a process plan is generated that shows the cycle time of each task.

To define human actions a motion capture device, Microsoft Kinect, is used to get accurate and realistic human motions for the given tasks. It integrates the mechanics of gaming technology into the DT environment to have higher degree of connectedness between human and the software. The human-operator performed the assigned tasks in front of the camera. The motions are real-time connected to the DT and each manual task was recorded and integrated with the robotic tasks.

General purpose robot controller can be used to run the simulation however accurate movements can be generated by using manufacturer specific robot controller. Robot controller for UR5 robot is integrated with the DT. The robot controller translates robot movements into robot understandable language.

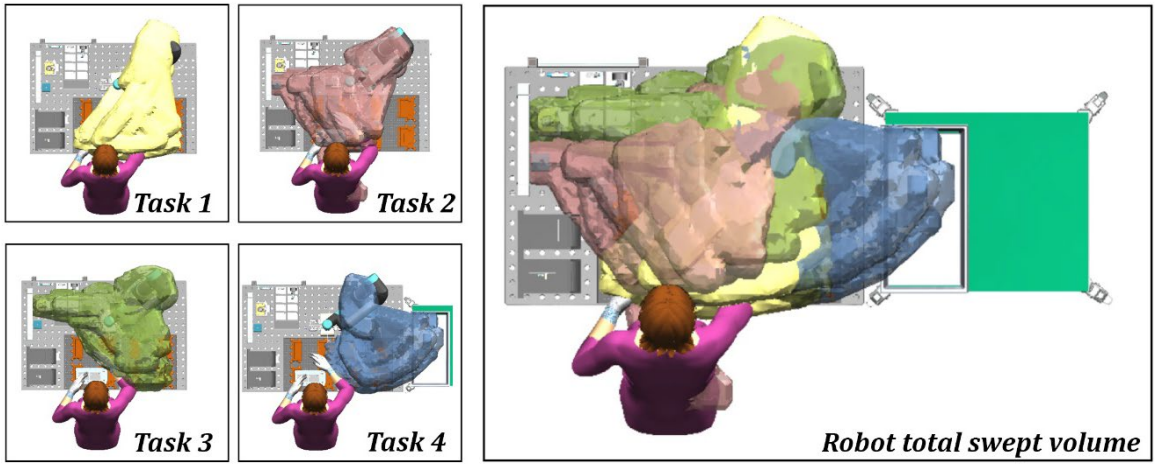

Figure 12: Robot swept volume.

After an optimal design for robot and human tasks is achieved, moving towards system integration, the robot program was tested in the real robot. The robot program was generated from the DT environment and was downloaded into the real robot. A live connection was developed between the simulation robot and the real robot. The physical robot followed the defined movements in the robot program but was placed in an empty space however, the virtual robot environment was equipped with hardware and virtual human. The collisions identified points where an optimization in layout was needed (Figure 13).

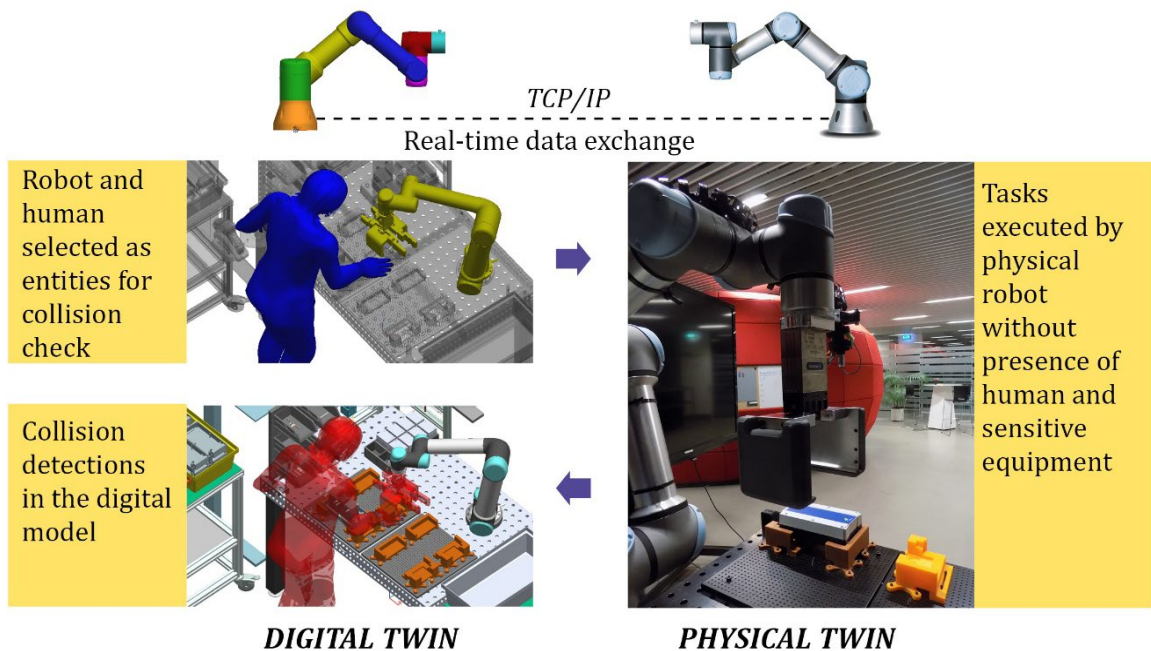

Figure 13: Online optimization of robot trajectories.

### 4.4. Building the physical twin

After generating the necessary design information, the next step is to generate bill of materials from the digital model. For each robot task, a gripping device is needed to pick the parts securely and move to the point of action or delivery. A part presentation strategy is needed defining how the parts are presented to the robot for picking the right part from the desired location each time and fixtures for the robot to hold and accurately assemble the parts at desired locations.

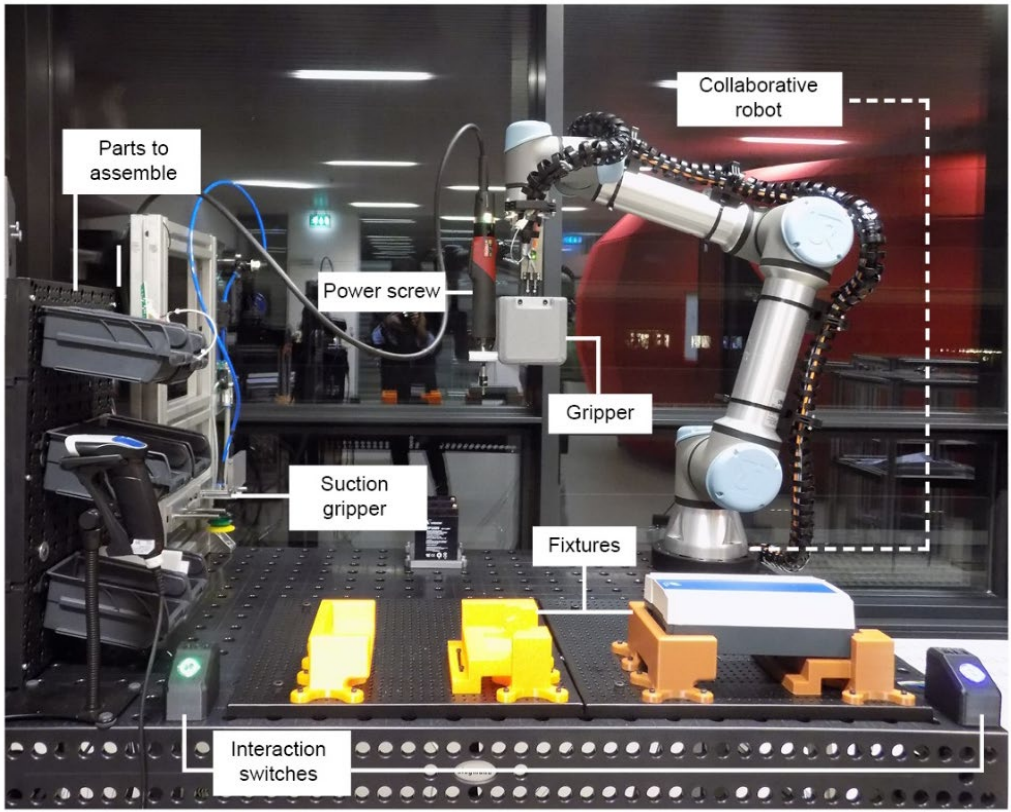

Figure 14: Physical twin developed in the case.

The physical twin is developed as a modular fixturing table (Figure 14). The fixture required to hold the product during HRC assembly are made with 3D printing. The fixture is needed both for the operator and the robotic tasks. Suction device is designed for gripping parts with no gripping features available. An electric screwdriver is another device that robot needs to handle. Mountings to hold these devices are designed with 3D printing to enable the robot to grasp the devices during operation. The screwdriver is connected to the robot controller that can be used to control the screwing operation. The utilization of the HRC system is shown in Figure 15 and is visible that both the robot and the human remain active during the cycle and it takes nearly 87 sec to complete the assembly.

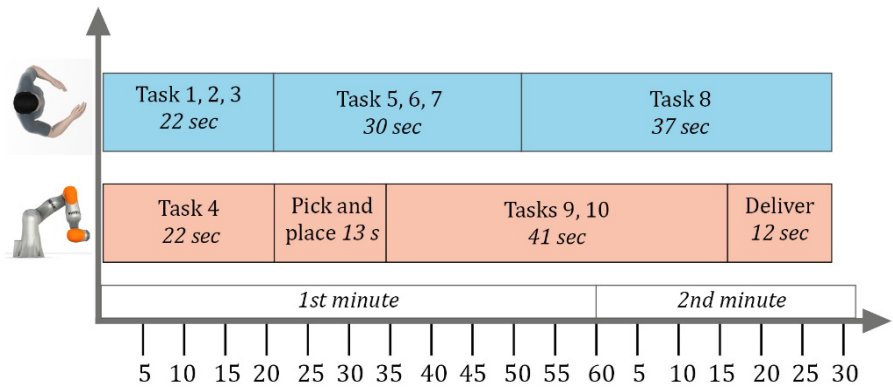

Figure 15: Utilization of the robot and the human operator.

### 4.5. Data logging

The cobot is connected with a cloud data repository through an IoT router. Three variables are identified to be logged that can affect system performance. The variables to log are: (1) how often the force torque sensor in robot arm gets activated. The force torque sensor enables the robot to alter its movement under a predefined force (10N in the case example); (2) the idle times (if any) when the robot is waiting for the operator to activate it to perform the task. There are two switches (one for each assembly fixture) available for the operator to activate the robot for each task; and (3) the number of completed assemblies. The logics are defined in the robot program as shown in Figure 16.

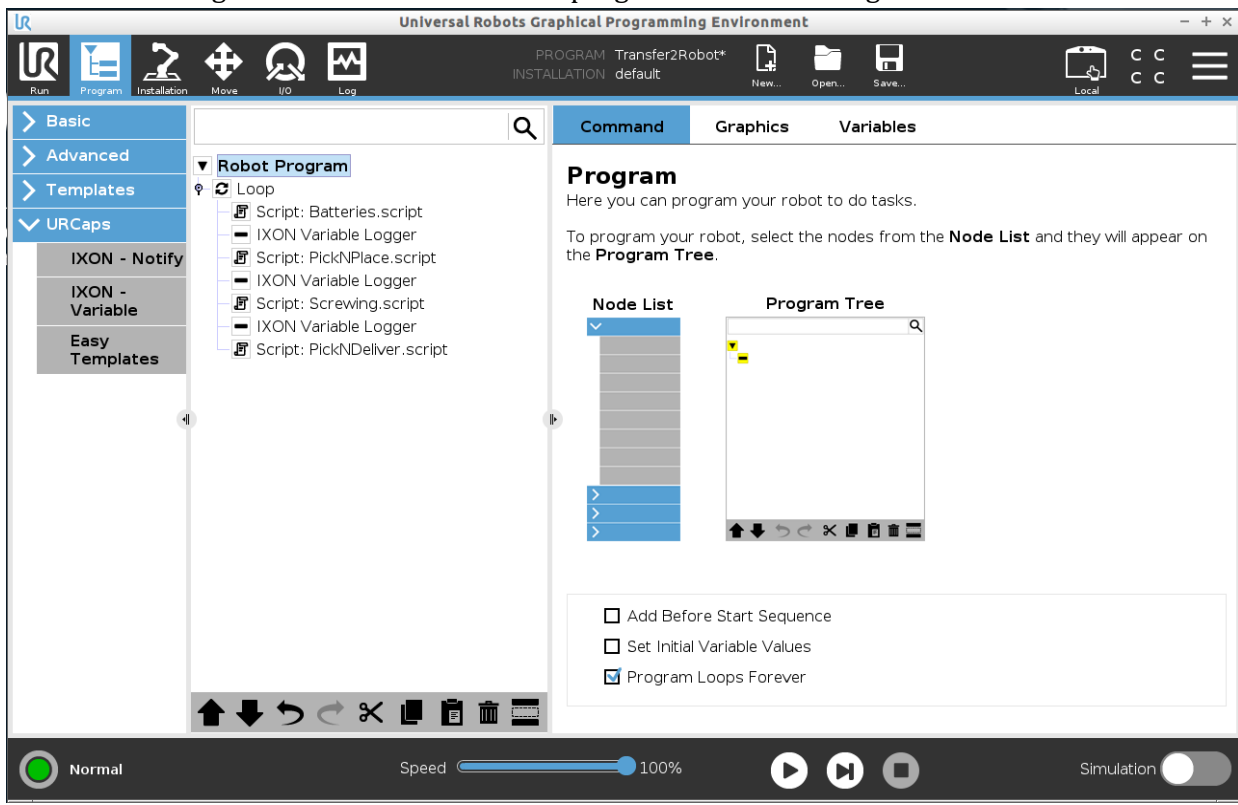

Figure 16: Data logging from the cobot.

A total of 30 assemblies were completed. The above-mentioned logs were recorded in the connected cloud. During completion of 30 assemblies, the robot and human collided 14 times. A significant waiting time (> 20 sec) was observed in total of 8 times and a variation in assembly cycle times was observed. The data was used in the digital twin environment and got a result of producing 326 assemblies during an 8 hours shift. To refine the process, the location of the fixtures was changed and the components available to the operator were replaced. The cobot trajectory was optimized by creating a virtual wall at the ending axis of the fixtures, therefore the robot will not move beyond fixtures. The simulation was tested and the program was generated and downloaded to the robot.

## 5. DISCUSSION AND FURTHER RESEARCH

HRC manufacturing system is discussed as a network of production elements that jointly form a unified whole to produce the final assembly. Robots, as a source of mainstream automation, are getting closer to the human-work space. The boundaries between humans and machine workspace are getting blurred. The interaction of humans and machines has been described as a complex phenomenon.

When it comes to operational phase, achieving a real-time updating of the virtual model according to the physical dynamics has opportunities but several challenges. Dynamic task distribution in a given product family is achieved with the given experiment. Easier robotic programming helped to avoid programming as a separate task. However, the adaptability of the robot in fraction of a seconds is needed to adapt to human movements. The conception of a digital twin is advantageous and need to be explored further. Embedding artificial intelligence into the data collected can make the system get self-learned and make decisions according to the past experience.

The following table summarizes the key learnings from this HRC in practice:

| Challenge | Solution and success factors |
| --- | --- |
| Layout optimization with digital twin | Layout optimization of a physical HRC system through a digital twin is fully achievable with available tools and techniques. The integration of a dynamic digital human model makes it easier to generate adaptive human-robot trajectories. |
| Dynamic task allocation | As the task allocation problem arises when both the human and the robot are able to accomplish a task. By defining the tasks' rating, it becomes possible for the simulation to allocate the tasks between human and robot depending upon their availability and score. |
| Robot control program | The robot program is one big advantage of using digital twins. The program is communicable in both directions between digital and physical twin. |
| Online collision optimization | By having an online linking between physical robot in an empty space and a virtual robot populated with virtual equipment in a virtual environment helped to make safe estimations of any possible collisions during operation |
| Human-robot interface | Several forms of human-robot interfaces can be developed using a DT. The purpose is to make human able to communicate intuitively with the robot e.g. hand gestures, and smart watch. |
| Enabling flexibility and high-mix low-volume production | The digital twin is used to have process balancing to take care of increased work in process. Since the key locations |

|  | are saved in the simulation, for any change in task execution, the robot needs to adapt to a new path and reach the defined location. Thereby, robot or human can temporarily bypass their original tasks to help each other in accomplishing the waiting tasks. |
|---|---|
| Human ergonomic analysis | Digital human models can be used to perform human tasks and evaluate the biomechanical loads, fatigue and stress levels. This is an interesting area to be explored further in relation to a digital twin |

As always, this kind of research has also several limitations. However, these limitations also offer several insights for further research, which is outlined in the following.

First of all, this research was done in a single case study setup. The case was chosen based on several criteria; however, more cases are needed. For this, we call for more research on use cases in the combination with man, machine and work in a digital-twin setup. Also, cultural influences might also have an impact on HRC combinations, which were not explicitly considered in this setup. For instance, mimics and gestures have a very different meaning in different cultural environments. If this is considered, HRC can be even more efficient. So similar studies in other countries, e.g. in Asia, are encouraged. Finally, also more longitudinal studies would offer interesting insights in the dynamics of HRC in a digital twin setup.

## 6. CONCLUSION

A human robot collaborative system is a complex system which can be managed with a probe, sense and respond approach. Its usefulness is based on its promise of being agile, adaptable and safe for human colleagues. This requires that the system is designed, validated and controlled with innovative approaches. The idea of a digital twin simulates the behavior of the system as a 'forward run' by creating time dependent virtual models of the physical system. With each change in production parameters new variables are simulated to predict a future behavior. The behavior can be visualized and results are assessed without the risk of any financial loss or human injury. This risk would be present in real production.

In the design phase the digital twin was used to design each component of the assembly system by considering its dynamics in relation to the rest of the system and its lifecycle. The dynamic simulation helped enable quantitative analysis of the feasibility or business vale of the proposed solution. The assessment tools also made it possible to choose the right equipment given the layout size constraints. With advances in information and communication technologies, the DT can continuously be evolved in real time offering greater usefulness at system level. The main contributions of this paper are as follows. (1) It presents the need and usefulness of digital twins in designing, developing and operating human-robot production systems. (2) A demonstrator is presented of an HRC digital twin based on an industrial case; Several results from design to operation of an HRC system are discussed.